\documentclass{eas}
\usepackage{graphicx}
\usepackage[utf8]{inputenc}

\begin{document}

\title{Photometric quality of Dome C \\ for the winter 2008 from ASTEP South} 
\author{Nicolas Crouzet}\address{University of Nice Sophia Antipolis, CNRS, Observatoire de la C\^ote d'Azur, B.P. 4229, 06304 Nice Cedex 4, France}
\author{Tristan Guillot}\sameaddress{1}
\author{Karim Agabi}\address{University of Nice Sophia Antipolis, CNRS, Observatoire de la Côte d'Azur, 06108 Nice Cedex 2, France}
\author{Yan Fante\"\i -Caujolle}\sameaddress{2}
\author{Francois Fressin}\address{Harvard-Smithsonian Center for Astrophysics, 60 Garden Street, Cambridge, MA 02138, US}
\author{Jean-Pierre Rivet}\sameaddress{1}
\author{Erick Bondoux}\sameaddress{2}\secondaddress{Concordia Station, Dome C, Antarctica}
\author{Zalpha Challita}\sameaddress{2,4}
\author{Lyu Abe}\sameaddress{2}
\author{Alain Blazit}\sameaddress{2}
\author{Serge Bonhomme}\sameaddress{1}
\author{Jean-Baptiste Daban}\sameaddress{2}
\author{Carole Gouvret}\sameaddress{2}
\author{Djamel M\'ekarnia}\sameaddress{2}
\author{Francois-Xavier Schmider}\sameaddress{2}
\author{Franck Valbousquet}\address{Optique et Vision, 6 bis avenue de l'Est\'erel, BP 69, 06162 Juan-Les-Pins, France}
\author{the ASTEP Team}

\begin{abstract}

ASTEP South is an Antarctic Search for Transiting ExoPlanets in the South pole field, from the Concordia station, Dome C, Antarctica. The instrument consists of a thermalized 10 cm refractor observing a fixed $3.88\,^{\circ}$~x~$3.88\,^{\circ}$ field of view to perform photometry of several thousand stars at visible wavelengths (700-900 nm). The first winter campaign in 2008 led to the retrieval of nearly 1600 hours of data. We derive the fraction of photometric nights by measuring the number of detectable stars in the field. The method is sensitive to the presence of small cirrus clouds which are invisible to the naked eye. The fraction of night-time for which at least 50\% of the stars are detected is 74\% from June to September 2008. Most of the lost time (18.5\% out of 26\%) is due to periods of bad weather conditions lasting for a few days ("white outs"). Extended periods of clear weather exist. For example, between July 10 and August 10, 2008, the total fraction of time (day+night) for which photometric observations were possible was 60\%. This confirms the very high quality of Dome C for nearly continuous photometric observations during the Antarctic winter. 

\end{abstract}

\maketitle


\section{Introduction}

The duty cycle in winter is a key parameter to evaluate the potential of Dome C for astronomical observations. The ASTEP project (Antarctica Search for Transiting ExoPlanets) aims to find extrasolar planets from Dome C and to qualify the site for photometry. The first campaign took place during the winter 2008 with the ASTEP South experiment. First we present the instrument. Then we evaluate the photometric fraction with two different analysis. Finally we present the duty cycle of ASTEP South for the whole campaign.

\section{The ASTEP South experiment}

The ASTEP project (\cite{Fressin2005}) is divided in two phases. The main ins\-trument, ASTEP400, is a 40 cm telescope under development. We are now testing it and the first campaign will take place during the winter 2010. In the meantime a small cheap instrument, ASTEP South, has been observing during the winter 2008 and is now at the middle of the second campaign.

ASTEP South consists of a 10 cm refractor, a front-illuminated 4kx4k pixel CCD camera, and a simple mount in a thermalized enclosure (\cite{Crouzet2009}). The refractor is a TeleVue NP101 and the camera is a ProLines series by Finger Lake Instumentation equipped with a KAF-16801E CCD by Kodak (for the choice of the camera see \cite{Crouzet2007}). Its quantum efficiency peaks at 63 \% with a mean of 50 \% in the spectral range 600-800 nm. The pixel size is 9 $\mu$m and the total CCD size is 3.7 cm. The pixel response non uniformity is around 0.5\%. Pixels are coded on 16 bits giving a dynamic range of 65535 ADU and the gain is 2.0 e-/ADU. We use a GM 8 equatorial mount from Losmandy. A thermalized enclosure is used to avoid temperature fluctuations: the sides of this enclosure are made with wood and polystyrene and a double glass window reduces temperature variations and its accompanying turbulence on the optical path. The windows are fixed together by a teflon part and separated by a 3 mm space filled with nitrogen to avoid vapour mist. The enclosure is thermalized to $-20^{\circ}$C and fans are used for air circulation. 

\begin{figure}[ht]
\centering
\resizebox{8cm}{!}{\includegraphics{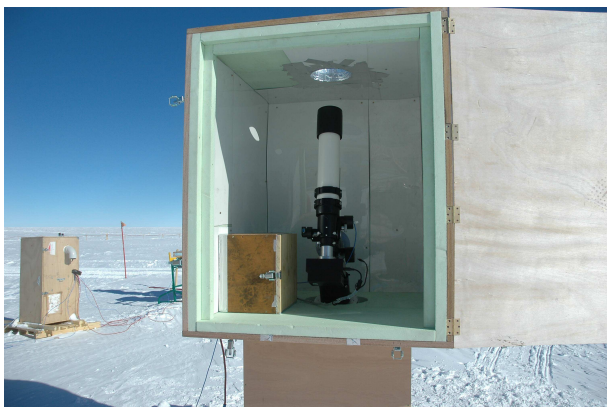}}
\caption{ASTEP South at Dome C, Antarctica, January 2008.}
\label{fig:ASTEPSouthDomeC}
\end{figure}

The ASTEP South instrument is shown at Dome C in figure~\ref{fig:ASTEPSouthDomeC}. In order to avoid as much as possible instrumental noises and in particular jitter noise, we chose a new observation strategy: the instrument is completely fixed and points towards the celestial South pole continuously. The observed field of view is $3.88\,^{\circ}$~x~$3.88\,^{\circ}$ leading to a pixel size of 3.41 arcsec on the sky. This field contains around 8000 stars up to Mv = 15. This observation setup leads to stars moving on the CCD from frame to frame and to an increase of the PSF (Point Spread Function) size in one direction depending on the exposure time.\\
Test observations made at the Calern observatory pointing towards the celestial North pole allowed us to choose an exposure time of 30 second and a PSF size of 2 pixel in FWHM, leading to only 2 saturated stars and to a limiting magnitude of 14. An analysis of the celestial South Pole field from the Guide Star Catalog with the same parameters leads to less than 10 \% of contaminated stars. The instrument was set up at the Concordia base in January-February 2008. The preliminary ana\-lysis of the 2008 campaign is presented here, focusing on the photometric quality of Dome C .

\section{Clear sky fraction at Dome C: a first estimate}

\cite{Mosser2007} estimated a clear sky fraction of 92 \% at Dome C for the winter 2006 observing the sky by naked eye several times a day. A more precise measurement is made with ASTEP South using the photometry of stars. From June 10\textsuperscript{th} to September 30\textsuperscript{th} 2008 we had 20.9 days of "white-out", ie days with a very cloudy sky often associated to storms and high temperatures (eg $-50^{\circ}$C). In this case stars are not visible at all. This gives a fraction of 18.5~\% of time unusable for observations. Another crucial parameter is the presence of high altitude clouds like cirrus. Although not visible by eye they absorb the star light and affect the photometry. No previous measurements have been made at Dome C regarding these clouds.

\begin{figure}[ht]
\centering
\includegraphics[width=8cm]{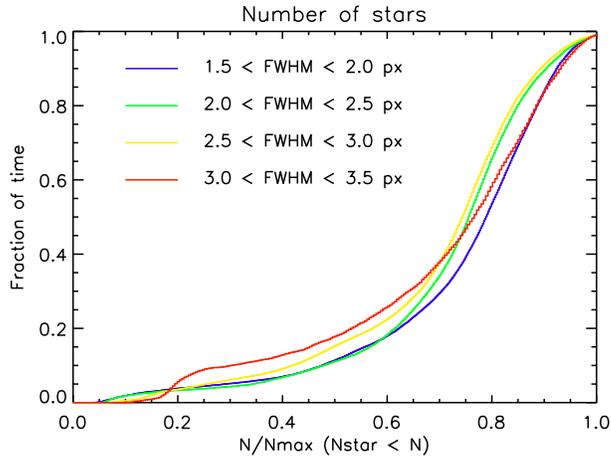}
\caption{Fraction of time for which we observe at least a given number of stars. The number of stars is normalized by the maximum number of stars expected for each full with half maximum.}
\label{fig:nbstars-cumulative}
\end{figure}

We derive the clear sky fraction with several methods. First, we consider that the sky is clear if we observe only half of the expected stars or less. This allows to take into account high altitude clouds. Periods affected by the Sun, typically few hours a day when the Sun is above -9 degrees, are excluded. The cumulative diagram figure~\ref{fig:nbstars-cumulative} shows the fraction of time with at least a given number of stars in our images. We show that i) at least 77 \% of the stars are visible for half of the time and ii) only half of the stars or less are visible for 15 \% of the time. Following our criterion this gives a clear sky fraction of 85 \%.  These fractions are derived from the periods where data were acquired, ie excluding most of the white-out periods during which the acquisitions are generally stopped. Considering all the white-out periods, we obtain a clear sky fraction of 74 \%.

\begin{figure}[ht]
\centering
\includegraphics[width=8cm]{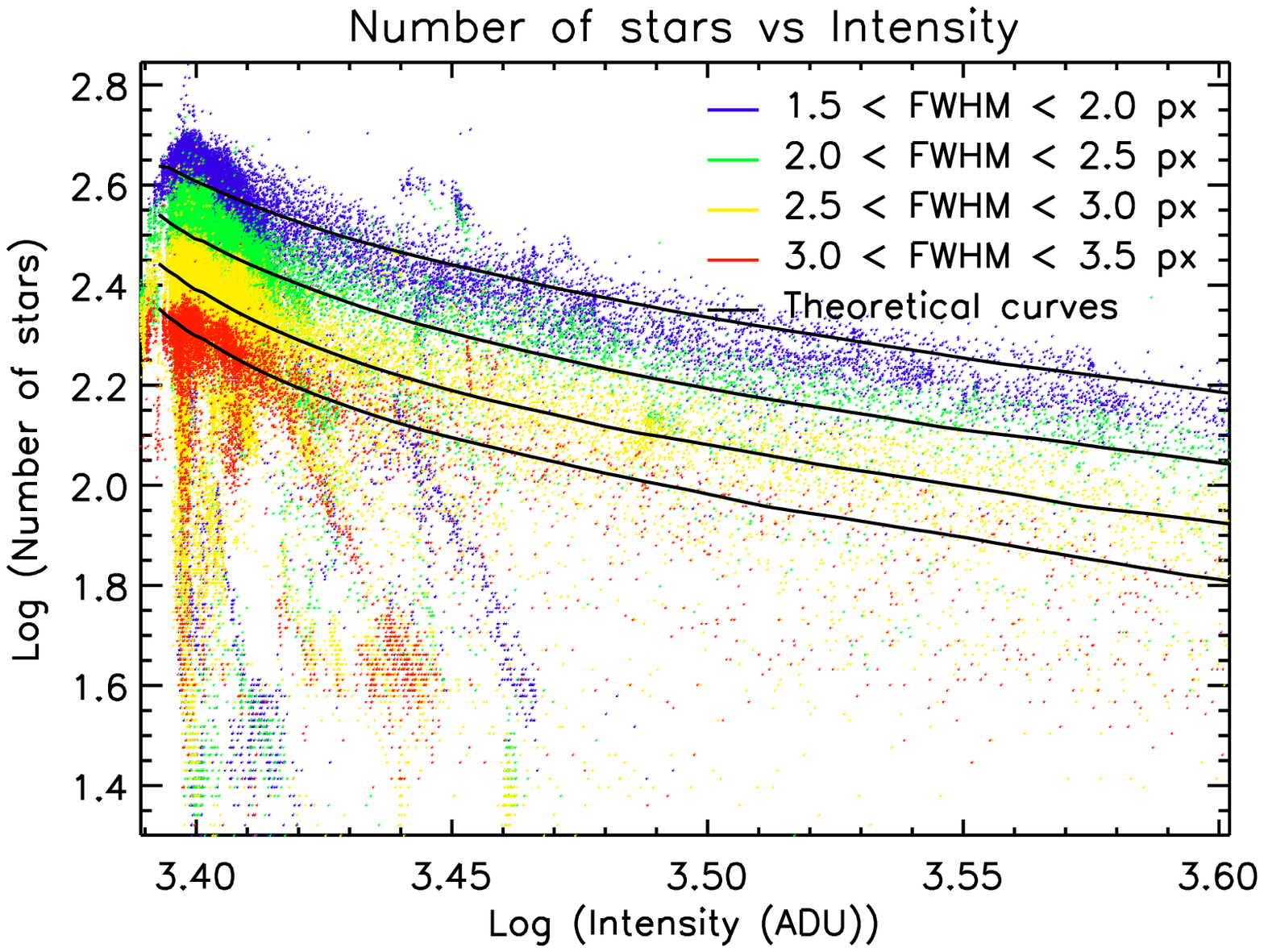}
\includegraphics[width=8cm]{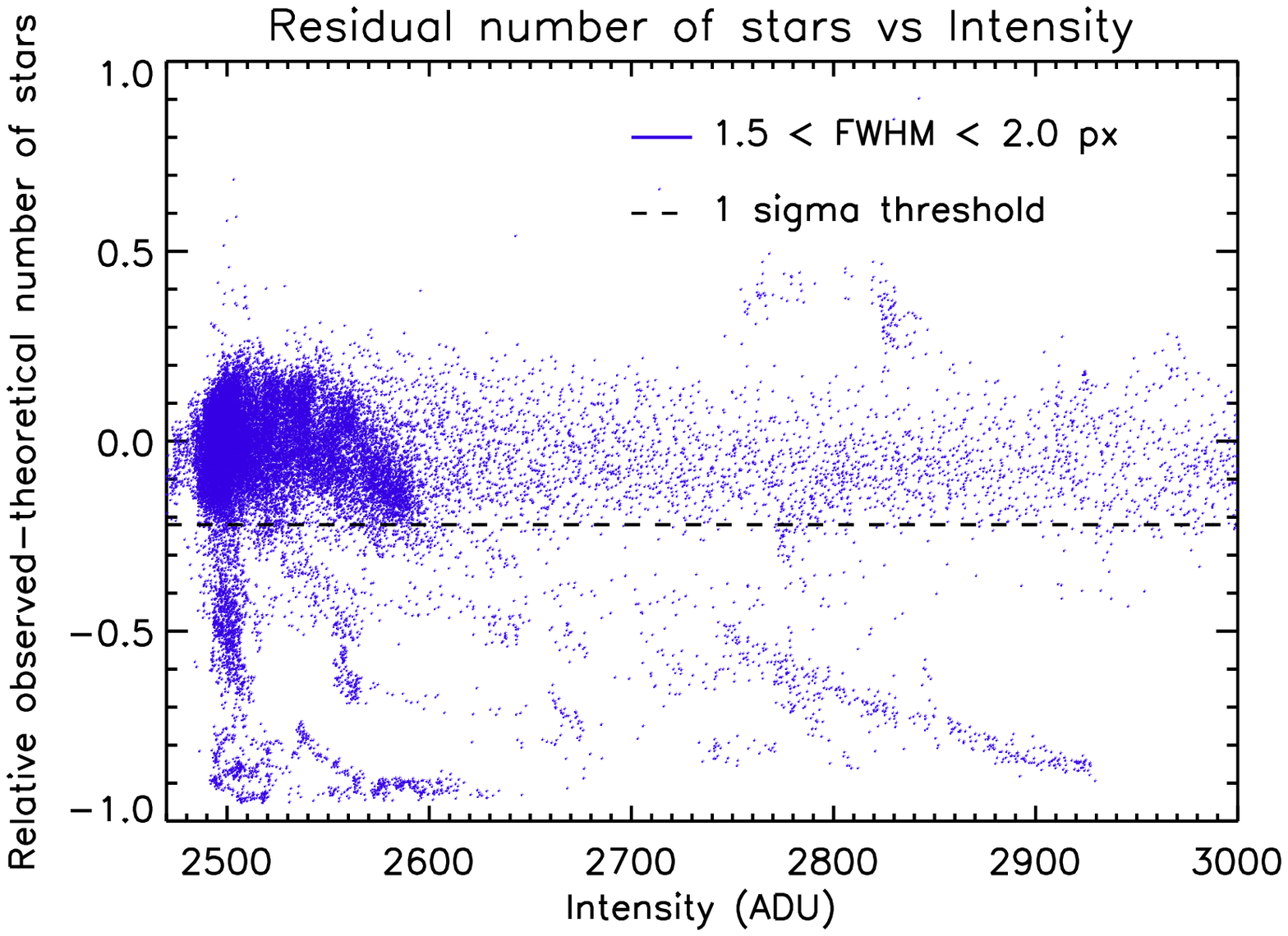}
\caption{Number of stars detected for a given background intensity for different FWHM. Measurements are in colors and theoretical curves in black (top). The difference between both is shown for one FWHM and allows to estimate the clear sky fraction (bottom).}
\label{fig:nbstars-intensity}
\end{figure}

Second we compare the expected to the actual number of stars detected for a given background intensity (figure~\ref{fig:nbstars-intensity}). Indeed a high background dilutes the faint stars into noise. As a increase of the PSF size will cause the same effect we separate different FWHM. Most of the points are spread around the theoretical curves but some measurements give a much lower number of stars than expected revealing the presence of clouds. The relative difference between the measured and theoretical values is used to estimate the fraction of clear sky. A threshold of 1 sigma of this distribution appear to separate well both regions and gives a clear sky fraction of 84.7 \%. Again considering all the white-out periods we obtain a clear sky fraction of 73.8 \%.

\section{Duty cycle of ASTEP South for the 2008 campaign}

We acquired nearly 1600 hours of data with ASTEP South for the 2008 campain. To evaluate the amount of photometric data the winter is divided in two minute periods. A period must contain at least one image with at least half of the expected number of stars to be considered as photometric. We do not consider seeing variations at the ground level which are important but very small above the 30 meter high boudary layer (\cite{Agabi2006}, \cite{Aristidi2009}). This gives a total of 1010 hours of photometric data. 

\begin{figure}[ht]
\centering
\includegraphics[width=8cm]{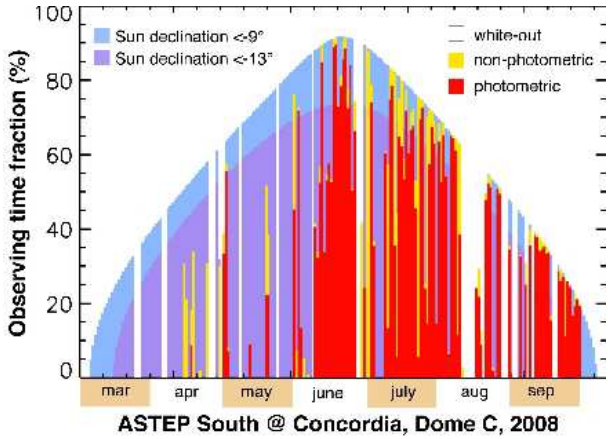}
\caption{Duty cycle for ASTEP South at Dome C for the 2008 campaign. Each bar represents one day. In blue is shown the Sun limit: in dark blue the fraction of time for which the Sun has no effect on the sky background (altitude $< -13^{\circ}$), and in light blue the fraction of time for which photometry is possible (altitude $< -9^{\circ}$). In yellow is the observing time fraction and in red the photometric time fraction of ASTEP South. White parts are the white-out periods, during which observations are not possible.}
\label{fig:DutyCycle}
\end{figure}

The duty cycle for the whole campaign of ASTEP South is represented in the histogram figure~\ref{fig:DutyCycle} in which each bar stands for one day. The limit due to the Sun, the observing time fraction and the photometric time fraction are shown as well as the white-out periods. Without considering the periods affected by the Sun, the ratio between the observing time plus white-out periods on one hand and photometric time on the other hand gives the photometric time fraction for the Dome C site. This results in a photometric fraction of 74 \% in agreement with the previous methods. As a comparison, the photometric fraction between 1991 and 1999 for the La Silla observatory in Chile is 62 \%. Moreover, the days of very bad weather are often grouped allowing long periods of continuous observations. For example we observed almost continuously during one month between July 9\textsuperscript{th} and August 8\textsuperscript{th}. Considering the photometric fraction and the hours lost because of the Sun, the total fraction of time usable for photometry for this one month period is 60 \%. In La Silla, taking also into accout the effect of the Sun, the mean photometric fraction for a one month period is 27~\% with a maximum of 45 \% in April 1997. This shows the very high quality of Dome C for continous observations and photometry during the Antarctic winter.


\section{Conclusion}

We have presented first results obtained from the ASTEP South 2008 campaign. These results confirm the high photometric quality of Dome C during the Antarctic winter, with a fraction of photometric night-time of 74\%. The possibility to observe nearly-continuously (with interruptions of a few hours around noon) during extended periods of time is favorable for the projects aimed at the detection and/or characterization of transiting planets. At the time of this writing, ASTEP South is in the middle of the 2009 winter season and functionning nominally. The next phase of the project, ASTEP 400 will consist in a pointable 40 cm Newton telescope to be installed at Concordia in 2010.

\section{Acknowledgements}

ASTEP has been funded by the Agence Nationale de la Recherche, the Institut des Sciences de l'Univers and the Programme National de Plan\'etologie. Operations at Concordia were made possible by the Institut Paul Emile Victor. ASTEP is led by the Observatoire de la C\^ote d'Azur and Universit\'e de Nice-Sophia Antipolis. Other participating institutes include the Observatoire de Haute Provence, DLR and the University of Exeter. NC acknowledges funding by the Observatoire de la C\^ote d'Azur and the R\'egion Provence Alpes C\^ote d'Azur.

\end{document}